\documentclass[10pt,a4paper]{article}

\usepackage{amscd}

\begin{document}

\title{Quantum invariance group of bosons and fermions}
\author{M. Arik, U. Kayserilioglu\\
       {\it Bo\~ gazi\c ci University, Physics Department}\\
       {\it Bebek 80815, Istanbul, Turkey} \\}
\date{\today}

\maketitle
\begin{abstract}
The particle algebras generated by the creation/annihilation
operators for bosons and for fermions are shown to possess quantum
invariance groups. These structures and their sub(quantum)groups
are investigated.
\end{abstract}

\newcommand{\beq}{\begin{equation}}
\newcommand{\eeq}{\end{equation}}
\newcommand{\bea}{\begin{eqnarray}}
\newcommand{\eea}{\end{eqnarray}}
\def\inbar{\,\vrule height1.5ex width.4pt depth0pt}
\def\IC{\relax\,\hbox{$\inbar\kern-.3em C$}}
\def\IR{\relax\,\hbox{\inbar\kern-.1em R}}
\def\FIO{$FIO(2d, \IR)\;$}
\def\BISp{$BISp(2d, \IR)\;$}

The concepts of bosons and fermions lie at the heart of
microscopic physics. They are described in terms of creation and
annihilation operators of the corresponding particle algebra: \bea
c_i c_j \mp c_j c_i & = & 0 \\
c_i c^*_j \mp c^*_j c_i & = & \delta_{ij}
\eea
where the upper sign is for the boson algebra $BA(d)$ and the lower sign is for the fermion algebra $FA(d)$.

It has been realized that quantum algebras play an important role
in the description of physical phenomena. Some classical physical
systems which are invariant under a classical Lie group, when
quantized, are invariant under a quantum group
\cite{fadeev,drinfeld,jimbo,woronowicz}. Thus quantum groups are
important in carrying over certain classical properties to the
quantum world.

Traditionally the boson algebra has the symmetry group $ISp(2d, \IR)$, the inhomogeneous symplectic group,
which transforms creation and annihilation operators as:
\beq
c_i \rightarrow \alpha_{ij} c_j + \beta_{ij} c^*_j + \gamma_i \quad .
\eeq
Here $\alpha_{ij}, \beta_{ij}, \gamma_i$ are complex numbers satisfying the constraints required by the group
$ISp(2d, \IR)$. One should note that this symmetry group is also the group of linear canonical transformations
of a classical dynamical system.
Similarly, the fermion algebra has the symmetry group $O(2d)$ with the transformation law:
\beq
c_i \rightarrow \alpha_{ij} c_j + \beta_{ij} c^*_j \quad .
\eeq
however, unlike its bosonic counterpart this algebra is not inhomogeneous.

Is there an analogue of the inhomogeneous invariance group for
fermions? It has already been shown in \cite{agy} that it exists
and is a quantum group for $d=1$. In this paper, we generalize the
fermion case to d-flavours and show that \FIO, the fermionic
inhomogeneous orthogonal quantum group is the invariance "group" of
$d$ fermions. If the same method is applied to bosons, the
resulting algebra is again a quantum group which we name \BISp,
the bosonic inhomogeneous symplectic quantum group. This quantum
group contains $ISp(2d)$ as a subgroup.

We first introduce the quantum groups \FIO and \BISp. Then we investigate their sub(quantum)groups. We will
also define $FIO(2d - 1, \IR)$, the fermionic inhomogeneous quantum orthogonal group in odd number of dimensions.

A general transformation of a particle algebra can be described in the following way:
\beq
\left(
\begin{array}{c}
c' \\
{c^{*}}' \\
1
\end{array}
\right)
=
\left(
\begin{array}{ccc}
\alpha & \beta & \gamma \\
\beta^* & \alpha^* & \gamma^* \\
0 & 0 & 1
\end{array}
\right)
\dot{\otimes}
\left(
\begin{array}{c}
c \\
c^* \\
1
\end{array}
\right)
\eeq
where $c, c^*, \gamma, \gamma^*$ are column matrices and $\alpha, \beta, \alpha^*, \beta^*$ are $d\times d$ matrices.
Thus, in index notation the transformation is given by:
\bea
c'_i &=& \alpha_{ij} \otimes c_j + \beta_{ij} \otimes c^*_j + \gamma_i \otimes 1 \quad , \\
{c^{*}}'_i &=& \alpha^*_{ij} \otimes c^*_j + \beta^*_{ij} \otimes c_j + \gamma^*_i \otimes 1 \quad .
\eea

Given this transformation, we look for an algebra $\mathcal{A}$ generated by these matrix elements such that
the particle algebra remains invariant. In such an algebra, successive transformations, the identity transformation
and inverse transformations should be meaningful hence this algebra should be a Hopf algebra.

Thus, we first write the transformation matrix in the above equation in the following way:
\beq
M =
\left(
\begin{array}{cc|c}
\alpha & \beta & \gamma \\
\beta^* & \alpha^* & \gamma^* \\
\hline
0 & 0 & 1
\end{array}
\right)
 =
\left(
\begin{array}{c|c}
A & \Gamma \\
\hline
0 & 1
\end{array}
\right) \quad .
\eeq
The coproduct, counit and coinverse defined as:
\bea
\Delta(M) & = & M \dot{\otimes} M \\
\epsilon(M) & = & I \\
S(M) & = & M^{-1} \eea should satisfy Hopf algebra axioms. We
assume that $\alpha_{ij}, \beta_{ij}, \gamma_i$ belong to a
possibly noncommutative algebra on which a hermitian conjugation
denoted by $*$ is defined. Motivated by \cite{agy} we also assume
that the matrix elements of $A$ form a commutative algebra. Then,
the inverse of the matrix $M$ can be defined by: \beq M^{-1} =
\left(
\begin{array}{cc}
A^{-1} & -A^{-1} \Gamma \\
0 & 1
\end{array}
\right)
\eeq
where $A^{-1}$ is defined in the standard way since matrix elements of $A$ are commutative.

Applying the invariance of bosonic/fermionic particle algebra and Hopf algebra relations to the matrix
elements of $M$, we get:
\bea
\gamma_i \gamma^*_j \mp \gamma^*_j \gamma_i &=& \delta_{ij} - \alpha_{ik}\alpha^*_{jk} \pm \beta_{ik} \beta^*_{jk} \\ \label{rel1}
\gamma_i \gamma_j \mp \gamma_j \gamma_i &=& \pm \beta_{ik} \alpha_{jk} - \alpha_{ik} \beta_{jk} \\ \label{rel2}
\eea where the upper signs are for \BISp and the lower signs for
\FIO. The other commutation relations are trivial in the sense
that the set $\alpha_{ij}, \beta_{i,j}, \alpha^*_{ij},
\beta^*_{ij}$ commute with each other whereas they
commute/anticommute with the set $\gamma_i$, $\gamma^*_i$ for
boson/fermions.

One question is what (quantum)subgroups do \FIO and \BISp have.
The subgroups are obtained by imposing additional relations on
the matrix elements of $M$. While searching for
sub(quantum)groups, we would also like to find new (quantum)groups
allowed by contractions \cite{inonu}. In order to do this, we
replace $\gamma_i$  by $\gamma_i / \sqrt{\hbar}$ so that we may
consider the case $\hbar \rightarrow 0$.

Thus on the algebra:
\bea
{[\gamma_i, \gamma^*_j]}_{\mp} &=& \hbar(\delta_{ij} - \alpha_{ik}\alpha^*_{jk} \pm \beta_{ik} \beta^*_{jk}) \\
{[\gamma_i, \gamma_j]}_{\mp} &=& \hbar(\pm \beta_{ik} \alpha_{jk} - \alpha_{ik} \beta_{jk})
\eea
we apply the relations:
\begin{enumerate}
\renewcommand{\labelenumi}{\bf(\alph{enumi})}
\item $\delta_{ij} - \alpha_{ik}\alpha^*_{jk} \pm \beta_{ik} \beta^*_{jk} = \pm \beta_{ik} \alpha_{jk} - \alpha_{ik} \beta_{jk} = 0$
\item $\gamma_i = 0$
\item $\beta_{ij} = 0$
\item $\alpha_{ij} = 0$
\item $\hbar \rightarrow 0$
\end{enumerate}
to get the sub(quantum)group diagram:
\[
\begin{CD}
{FIO(2d, \IR)} @>{\bf (a)}>> {GrIO(2d, \IR)} @>{\bf (b)}>> {O(2d, \IR)} \\
@V{\bf (c)}VV @V{\bf (c)}VV @V{\bf (c)}VV \\
{FIU(d)}  @>{\bf (a)}>> {GrIU(d)} @>{\bf (b)}>> {U(d)} \\
@V{\bf (d)}VV \\
{FA(d)}
\end{CD}
\]
for the fermionic case and the diagram:
\[
\begin{CD}
{BISp(2d, \IR)} @>{\bf (a)}>> {ISp(2d, \IR)} @>{\bf (b)}>> {Sp(2d, \IR)} \\
@V{\bf (c)}VV @V{\bf (c)}VV @V{\bf (c)}VV \\
{BIU(d)} @>{\bf (a)}>> {IU(d)} @>{\bf (b)}>> {U(d)} \\
@V{\bf (d)}VV \\
{BA(d)}
\end{CD}
\]
for the bosonic case.

For the contraction {\bf (e)} we get the tables:
\[
\begin{CD}
{FIO(2d, \IR)} @>{\bf (e)}>> {GrIGL(2d, \IR)} \\
@V{\bf (c)}VV @V{\bf (c)}VV \\
{FIU(d)} @>{\bf (e)}>> {GrIGL(d, \IC)} \\
@V{\bf (d)}VV @V{\bf (d)}VV \\
{FA(d)} @>{\bf (e)}>> {Gr(d, \IC)}
\end{CD}
\]
for the fermionic case and the diagram:
\[
\begin{CD}
{BISp(2d, \IR)} @>{\bf (e)}>> {IGL(2d, \IR)} \\
@V{\bf (c)}VV @V{\bf (c)}VV \\
{BIU(d)} @>{\bf (e)}>> {IGL(d, \IC)} \\
@V{\bf (d)}VV @V{\bf (d)}VV \\
{BA(d)} @>{\bf (e)}>> {\IC^d}
\end{CD}
\]
for the bosonic case.

In the above tables, $GrI$ stands for grassmanian inhomogeneous with the meaning that the translation
parameters are grassmanian. These can be considered to be inhomogeneous supergroups.

In our approach, for the fermionic case, $\alpha_{ij}$,
$\beta_{ij}$, $\alpha^*_{ij}$, $\beta^*_{ij}$ anticommute with
$\gamma_i$, $\gamma^*_i$ and the matrices $M$ are multiplied with
the standard tensor product. It can be shown that $\alpha_{ij}$,
$\beta_{ij}$, $\alpha^*_{ij}$, $\beta^*_{ij}$ can be taken to
commute with $\gamma_i$, $\gamma^*_i$ provided that the matrices
$M$ are multiplied with a braided \cite{majid} tensor product, eg.
\beq (A \otimes B)(C \otimes D) = - AC \otimes BD \eeq whenever
$B$ and $C$ are both fermionic. This corresponds to the usual
superalgebra approach.

To construct $FIO(d, \IR)$ for odd $d$, we perform a similarity
transformation
\[
M \rightarrow UMU^{-1}
\]
using the unitary matrix:
\beq U = \left(
\begin{tabular}{cc|c}
$\frac{1}{\sqrt{2}}$ & $\frac{1}{\sqrt{2}}$ & $0$ \\
$\frac{-i}{\sqrt{2}}$ & $\frac{i}{\sqrt{2}}$ & $0$ \\
\hline
$0$ & $0$ & $1$
\end{tabular}
\right)
\eeq
to put $M$ in real form. Then it is found that for
\FIO (\ref{rel1}) and (\ref{rel2}) are replaced by: \beq [\Gamma_i
, \Gamma_j]_+ = \delta_{ij} - A_{ik} A_{jk} \quad\quad, i, j =
1, 2, \ldots , 2d . \eeq By extending the range of the indices to
odd-dimension it is possible to define $FIO(d, \IR)$ also for odd
$d$.

Finally, we would like to remark that the widely used field
theoretical generalization achieved by extending the discrete
indices $i, j, k$ to continuous variables together with a
replacement of the Kronecker deltas to Dirac delta functions is
also applicable to the quantum groups we obtained. We believe that
the establishment of quantum groups in field theory will be
helpful in generalizing methods of quantization. These approaches
hopefully will yield a more consistent approach to interacting
field theory.

\end{document}